# Receive signal path design for Active phased array radars

Mohit Kumar, Dileep (Coreel Technology), K.Sreenivasulu, D.Seshagiri, Durga Srinivas and S.Narasimhan
Electronics and Radar Development Establishment (LRDE)
CV Raman Nagar, Bangalore-560093, INDIA
mohit.kumar@lrde.drdo.in

*Abstract*
　　*Modern Active Phased array Radar systems with large number of T/R modules, multi channel receiver down converters and distributed power distribution networks leads to design and analysis of the receive signal path more complex. In this paper receive signal path design of a typical 1000 T/R modules based fully distributed active phased array radar is discussed in details including the gain, Spurious Free Dynamic Range (SFDR) requirements at different levels. The techniques for optimization of SFDR and system performance also described along with the Systemvue model for receive path calculations is presented.*

*Key words*: Active Phased array Radar, SFDR, T/R modules, Receivers.

## I Introduction:

Designers of signal receiver systems often need to perform cascaded chain analysis of system performance from the antenna all the way to the ADC. In optimizing high dynamic range digital radar receivers, some non-straight forward trade-offs in the design considerations of the RF front-end should be taken into account. The designer should pay more attention in particular to the noise figure degradation and SFDR degradation and the linearity of every non-linear component in the receiver chain [2]. The SFDR degradation is directly linked to low dynamic range and noise figure degradation leads to range loss for targets.

SFDR is measure of the linearity of amplifiers, gain blocks, mixers, and other RF components. The second and third-order intercept points (IP2 and IP3) are figures of merit for these specifications and allow distortion products to be computed for various signal amplitudes. Two tone SFDR is measured by applying two spectrally pure sine waves to the device under test at frequencies f1 and f2, usually relatively close together. The amplitude of each tone is set slightly more than 6 dB below full-scale so that the amplifier does not clip when the two tones add in-phase. The second-order products fall at frequencies which can be removed by filters. However, the third-order products 2f2 – f1 and 2f1 – f2 are close to the original signals and are more difficult to filter.

Defining the third order Intercept point of a system is a way of characterizing a systems third order distortions and ultimately determining the high end of the system's SFDR. The third order spurs increase by 3 dB for every 1 dB increase in the fundamental frequency. A plot of the fundamental frequency's output power vs. input power would show a slope of 1:1, until we started to reach the systems non-linear region. A plot of the third order frequencies would show a slope of 3:1 in the linear region of the system. If you extend the linear portions of these two plots on the same graph, they would eventually intersect. This point where they intersect is called the Third Order Intercept point (IP3). The IP3 can be referenced to the input or the output. The input IP3 (IIP3) is the input power at the intercept point. The output IP3 (OIP3) is the output power at the intercept point. The IIP3 equals the OIP3, minus the system gain. This is shown in fig 1.

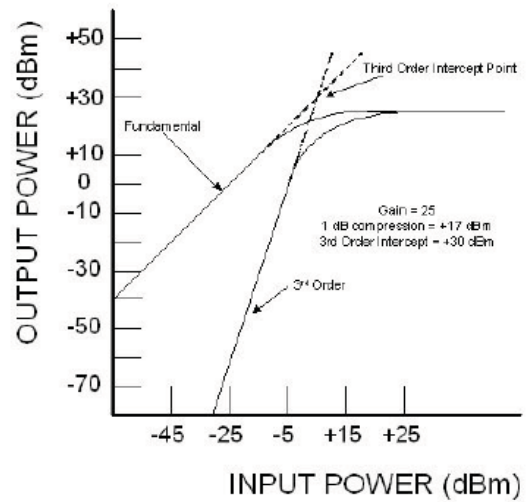

Fig: 1. Third Order intercept point(Log-log scale).

Once the IP3 is known, SFDR is calculated using the following equation: SFDR = 2/3 (IIP3- Noise Floor). The IIP3 for a cascade of stages with Gains $G_A, G_B, G_C$ respectively and IIP3 as $IIP3_A, IIP3_B$ and $IIP3_C$ is given by:
$1/IIP3^2_{Cas} = 1/IIP3_A^2 + G_A^2/IIP3_B^2 + G_A^2 G_B^2/IIP3_C^2$.

Noise figure (NF) is a measure of degradation of the signal-to-noise ratio (SNR), caused by components in RF signal chain. It is the difference in dB between the noise output of the actual receiver to the noise output of an "ideal" receiver with the same overall gain and bandwidth when the receivers are connected to matched sources at the standard noise temperature $T_0$ (usually 290 K).
NF = $10\log(SNR_{IN}/SNR_{OUT})$ = $SNR_{IN,dB} - SNR_{OUT,dB}$.
If several devices are cascaded, the total Noise Factor F is:
F = F1+F2/G1+F3/G1G2 + …. + Fn/G1G2…Gn-1

Where Fi is the noise factor of i[th] device and Gi is the gain of that device.

## II Design of Receiver chain:

Taking into account the above considerations of Noise figure and SFDR, it is relatively straightforward to design a receiver with low Noise figure by using an LNA (Low Noise Amplifier) in the front-end, so that the overall noise figure of the receiver is dominated by LNA noise figure alone [1] and the other terms in the cascaded noise figure equation becomes small.

To optimize cascaded SFDR, we need to either decrease the gain, which is usually not the choice given, or increase the IIP3 of the devices. We have to choose devices with higher IIP3 points. The last amplifier in the chain to have the highest IIP3 as it sees the highest gain in the chain. Based on these principles, given below is a sample receiver chain diagram on which we have carried out extensive simulations and theoretical calculations to optimize the design:

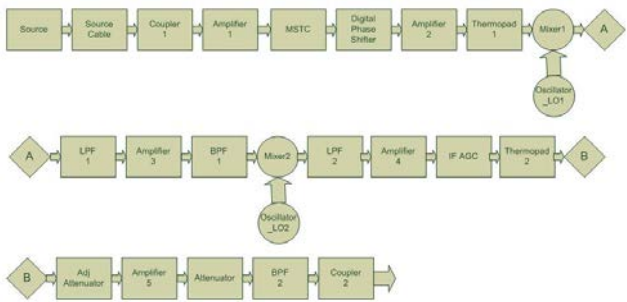

Fig: 2. A high performance Receiver design

The above chain has been designed to meet gain requirement of 42 dB. The amplifier 1 is an LNA with gain of 18dB and less than a dB noise figure. This takes care of the overall noise figure of the system. The amplifier 2, 3, 4 and 5 have gain of 18, 24, 20 and 20 dB respectively. These amplifiers have high IIP3 points. The SFDR obtained was of the order of 70 dB. This is an S-Band receiver with fixed LO2 of 540 MHz and switching LO1. The final IF is 60 MHz. The thermopads have been used to compensate gain variation with temperature. The RF simulation of the above chain is carried out in Agilent system Vue simulator.

## III Simulation using SystemVue:

The simulation has been carried out for the receiver chain in fig 2 for frequency in range 3.1-3.5GHz, power levels from -122 to -32 dBm and and operating temperature varies from -40 degree to +85 degC. RF signal is down converted into 60MHz using a two stage down conversion. The parameters assigned are as per the values obtained from datasheets and .s2p files provided by the vendors for respective components. Worst case analysis (maximum gain condition due to process) is also carried out at +25 degree at RF input frequencies 3100MHz, 3300MHz and 3500MHZ at a power level of -32dBm with Adj attenuator kept at 0dB.

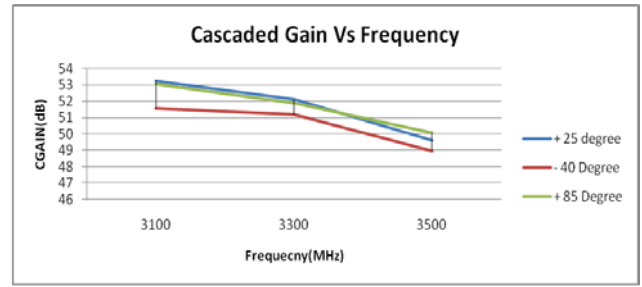

Fig: 3. Gain v/s Frequency over temp graph

The above figure shows gain variation over frequency and temperature. Gain variation over temp is of the order of 2dB which can be compensated by proper use of thermopads. However gain variation over frequency is of the order of 4 dB, which is compensated by using gain calibration tables for different frequencies which can be used to program Adj attenuators at different frequencies. Here the input power is at -32dBm.

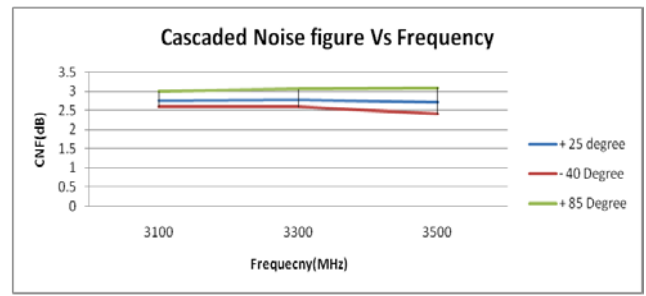

Fig: 4. Noise Figure v/s Frequency graph

The above figure gives variation of noise figure over frequency and over temperature. The noise figure is high at higher temperature as the thermal noise is higher with temp. The noise figure decreases slightly with frequency but largely remains constant over frequency. Here the input power is at -32dBm.

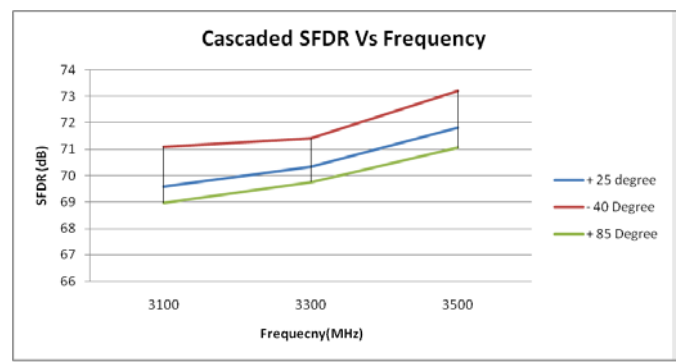

Fig: 5. Cascaded SFDR v/s Frequency graph

The above figure gives variation of SFDR over frequency and over temperature. As observed, the SFDR is lower for higher temp as the noise figure is poor at higher temp and vice versa. Since the noise figure is slightly good for higher frequencies, the SFDR is also higher as seen in above figure. Here, the RF input power and interferer power levels are both at -32dBm and the interferer is 1

MHz away from the main signal. This ensures that the third order intermods are within the passband (5MHz).

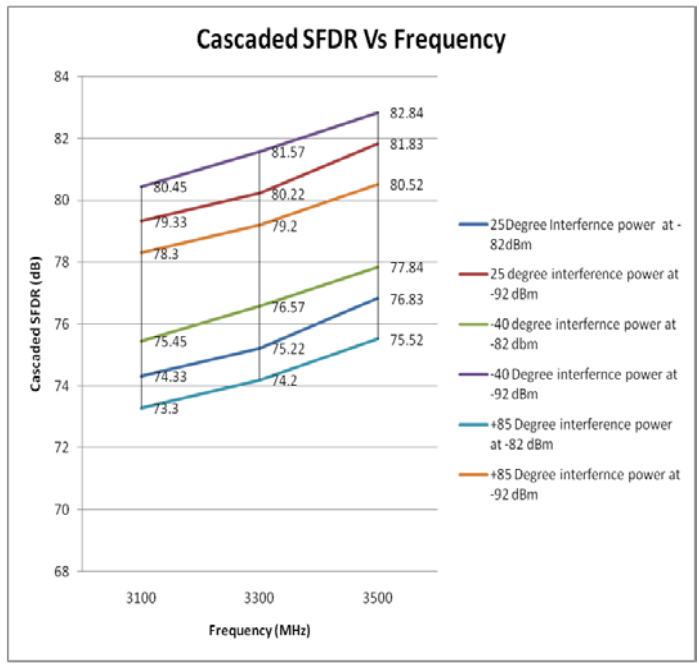

Fig: 6. Cascaded SFDR v/s Frequency with different power levels of interferer.

This above figure shows the same variation of SFDR with frequency and temp but also with different interferer power levels. For all temperatures, the two interferer power levels of -92dBm and -82dBm has been considered with the main signal still at -32dBm. The SFDR is better off with a lower interfering signal but it is not directly translating very highly to SFDR as the main signal is at -32dBm.

### V Conclusion:

The Radar receiver optimization problem have been studied analytically, and via simulation. The various parameters of Receiver like dynamic range, noise figure and Gain over frequency and temp range have been simulated and shown here and schemes of optimizing these have been discussed. One of the design of an active array radar receiver has been taken up and the optimization on that architecture were discussed and the results were elaborated

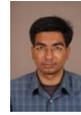
**Mohit Kumar** born on 7th October 1980 obtained B.Tech degree in Electronics and Communication from NIT, Jalandhar in 2002. He has completed his MTech from IIT Delhi in Communication Engg. in 2010. He is currently working as scientist at Electronics and Radar Development Establishment (LRDE), Bengaluru. Area of specialization is in Digital Radar Receiver design.

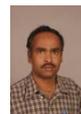
**K Sreenivasulu** received his Diploma in Electronics and Communication Engineering from S.V. Government Polytechnic, Tirupati, Andhra Pradesh in the year 1987. He received his B.Tech degree in Electronics and Communication Engineering from Jawaharlal Nehru Technological University, Hyderabad in the year 1995. He received M.E. degree in Micro Electronics Systems from Indian Institute Of Science, Bangalore in 2004. He started his professional career as Electronic Assistant in Civil Aviation Department where he worked from 1990 to 1995. Since 1996 he has been working as Scientist in Electronics and Radar Development Establishment (LRDE), Bangalore. His area of work has been design and development of RF and Microwave sub-systems, Digital Radar system, Beam Steering Controller for Active Aperture Array Radars. His interests include VLSI Systems and Programmable Controllers.

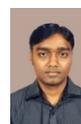
**G DurgaSrinivas** born on 2nd June 1986 obtained B.Tech degree in Electronics and Communication from NIT Warangal in 2008. He is currently working as a scientist at Electronics and Radar Development Establishment (LRDE), Bengaluru. Area of specialization is in Digital Receiver systems.